\documentclass[preprint2]{aastex}

\lefthead{Virgo Cluster Principal Axis}
\righthead{West and Blakeslee}

\begin{document}

\title{The Principal Axis of the Virgo Cluster}

\author{Michael J. West}
\affil{Department of Physics \& Astronomy, University of Hawaii at Hilo,
    Hilo, HI 96720}

\and

\author{John P. Blakeslee\altaffilmark{1}} 
\affil{Department of Physics, University of Durham, Durham, DH1 3LE, England}
\authoremail{J.P.Blakeslee@durham.ac.uk}

\altaffiltext{1}{Present address: Department of Physics \& Astronomy,
    Johns Hopkins University, Baltimore, MD 21218}

\begin{abstract}

Using accurate distances to individual Virgo cluster galaxies
obtained by the method of Surface Brightness Fluctuations, we show 
that Virgo's brightest ellipticals have a remarkably collinear 
arrangement in three dimensions.  This axis, which is inclined 
by $\sim 10 - 15^{\circ}$ from the line of sight, can be traced to even 
larger scales where it appears to join a filamentary bridge of galaxies 
connecting Virgo to the rich cluster Abell 1367.  The orientations of 
individual Virgo ellipticals also show some tendency to be aligned with 
the cluster axis, as does the jet of the supergiant elliptical 
M87.  These results suggest that the formation of the Virgo 
cluster, and its brightest member galaxies, have been driven  
by infall of material along the Virgo-A1367 filament.

\end{abstract}


\section{Introduction}

The Virgo cluster, at a distance of approximately $15\,h^{-1}$ Mpc, 
is the nearest richly-populated cluster of galaxies and, consequently, 
one of the best studied.  
A number of authors have pointed out that Virgo's brightest elliptical 
galaxies have a remarkably linear arrangement, along a
projected position angle of roughly $110^{\circ}$ 
(measured North through East).  \citet{arp68}, for example, noted that 
``all the E galaxies in the northern half of the Virgo cluster fall on 
a line going through M87.''  Similarly, \citet{bing87} suggested that 
``The line connecting M87 and M84 appears as a fundamental axis of the 
cluster.''   This can be seen in Figure 1, which plots 
the distribution of probable member galaxies in the northern portion 
of the Virgo cluster, as seen on the plane of the sky.   
The distribution of Virgo dwarf elliptical galaxies also appears  
somewhat elongated in this direction \citep{bing99}, as does the distribution 
of hot X-ray emitting intracluster gas \citep{bohr94,schind99}.   
However, without accurate distances to individual galaxies, it is impossible 
to say for certain whether Virgo's apparent principal axis is a genuine 
three-dimensional structure, or merely an illusory chance alignment of 
galaxies.  Furthermore, three-dimensional information would allow one 
to measure the true shape and spatial orientation of this axis.

Here we present evidence that Virgo's bright elliptical galaxies 
trace a highly elongated, three-dimensional structure that is 
actually a small segment of a much larger filament passing through the 
heart of the Virgo cluster.

\section{Resolving the Virgo Cluster in Three Dimensions}

A variety of methods exist for measuring galaxy distances (see 
Jacoby et al. 1992 for a review).  One of the most powerful is the 
technique of Surface Brightness Fluctuations (hereafter SBF), 
whereby distances to galaxies are estimated from the ratio of the 
second and first moments of their stellar luminosity functions  
(Tonry \& Schneider 1988; see Blakeslee, Ajar \& Tonry 1999 for a 
recent review). This method works best for early-type galaxies,
although it has also been applied to the bulges of some spirals.
\citet{ferrarese00} have concluded that SBF is
the most accurate early-type galaxy distance indicator 
reaching cosmologically interesting distances.

\citet{tonry00} have recently published SBF distances for 300 nearby
galaxies that were observed as part of their $I$-band SBF Survey.  Based
on distances for 31 probable Virgo cluster members, they derived a mean
cluster distance modulus $m-M = 31.15 \pm 0.03$, corresponding to a
distance of $17 \pm 0.3$ Mpc.  Uncertainties in individual Virgo galaxy
distances are $\sim 1.5-2$ Mpc, which suggests that it should be possible
to resolve the Virgo cluster along the line of sight, at least partially.
Indeed, Tonry, Ajhar, Luppino (1990) attempted to do this a decade ago, but
they lacked a suitable calibration. They assumed that the $I$-band SBF
magnitude $\overline m_I$ was insensitive to stellar population variations
among ellipticals, as implied by the best population models at the time. 
As a result, they essentially found that bluer Virgo ellipticals were
systematically in front of redder ones.  
Following a decade of work,
the stellar population dependence of $\overline m_I$ has been empirically
well characterized and calibrated out \citep{tonry97, ferrarese00},
so that there is no longer any correlation of the distances
with stellar population parameters.  Moreover, the empirical calibration
now has strong support from theoretical modeling 
\citep{worthey94, liu00, blakeslee00}.

Table 1 lists data for all elliptical galaxies in the northern 
portion of the Virgo cluster with SBF distances from the
\citet{tonry00} survey.  Column (1) gives the galaxy name, 
column (2) lists the SBF distance of each galaxy from \citet{tonry00}, 
and column (3) gives the orientation of the galaxy major 
axis, taken from the Lyon-Meudon Extragalactic Database (LEDA).  
The positions and distance moduli of these galaxies are also indicated 
in Figure 2.  

Table 1 and Figure 2 suggest some tendency for those galaxies located 
in the western region of the cluster to be more distant than those 
on the eastern side.  This trend can be seen more clearly in Figure 3, 
which shows a clear correlation between galaxy right ascension and 
distance, with a systematic trend of increasing galaxy distance as 
one moves in the westerly direction.  Pearson (parametric) and Spearman 
(non-parametric) rank correlation tests both confirm that this trend is 
statistically significant, at the 97\% and 99\% confidence levels, 
respectively (the correlation becomes even stronger if NGC 4168 is 
also included).  We note that Neilsen \& Tsvetanov (2000) have recently 
measured independent SBF distances with the {\it Hubble Space Telescope} 
for 15 Virgo galaxies, including 10 ellipticals along this central
ridge line, and have observed the same trend in their data.

To measure the three-dimensional shape and orientation of this axis, 
we computed the moments of inertia of the system of galaxies in Table 1 
after converting their right ascensions, declinations and distances to 
supergalactic cartesian coordinates.  
Diagonalization of the inertia tensor yields the three eigenvalues 
corresponding to the principal moments of inertia, 
and the associated eigenvectors provide information on the orientation 
of the principal axis.

We find that the Virgo ellipticals have a remarkably collinear arrangement 
in three-dimensions, with an {\it rms} scatter of only $\sim 400$ kpc
about the principal inertial axis along its $\sim 8$ Mpc length between 
NGC 4387 and NGC 4660.  This axis is inclined at an angle of 
approximately $\sim 75^{\circ}-80^{\circ}$ with respect to the plane of 
the sky, i.e., close to the line of sight.  Presumably with smaller distance 
uncertainities, Virgo's principal axis would be found to be even narrower.

\section{Galaxy Orientations in Virgo}

Additional evidence of the special nature of the axis defined by 
Virgo's brightest ellipticals comes from the orientations of the 
galaxies themselves.  As Table 1 shows, the majority have projected 
major axis position angles between $100^{\circ}$ and $140^{\circ}$, 
quite similar to the $110^{\circ}$ 
projected orientation of the cluster principal axis.  
A Kolmogorov-Smirnov (KS) test indicates a probability of only $\sim 7\%$ 
that the galaxy position angles in Table 1 are randomly-oriented between 
0 and 180 degrees.  While one must be cautious not to overinterpret the 
statistical significance of results based on a small sample like this, 
nevertheless it is suggestive that the orientations of Virgo's large 
ellipticals may be somehow related to the direction of the cluster 
principal axis.   A similar alignment effect is seen for the 
brightest elliptical galaxies in the Coma cluster \citep{west98},
as well as other clusters \citep{bing82,porter91,west94} and references 
therein.
 
To test whether fainter Virgo ellipticals might also have preferred 
orientations with respect to the cluster major axis, we used 
the Virgo Photometry Catalogue (VPC) of \citet{yc98}, which provides 
data, including orientations, for 1180 galaxies in the core region of 
the cluster.  Because the VPC includes both Virgo members and unrelated 
galaxies along the line of sight, we cross-correlated it with the list 
of probable Virgo members from \citet{bing85} to produce a sample of 
108 Virgo ellipticals with measured major axis position angles.  
Application of the KS test to this sample shows no statistically 
significant tendency for the fainter ellipticals to have preferred 
orientations.  However, when the roundest galaxies $-$ whose position 
angles are most uncertain $-$ are eliminated by restricting the 
sample to only those with ellipticities greater than 0.2, then the 
KS test indicates that this subset of 69 galaxies has only a 
$\sim 4\%$ probability of being consistent with a randomly orientated 
population, with a median galaxy position angle of $107^{\circ}$.

\section{A Link Between Virgo and the A1367-Coma Supercluster?}

One of the most striking features of the large-scale distribution 
of galaxies is its filamentary appearance, with long, quasi-linear 
arrangements of galaxies that extend tens or perhaps even hundreds 
of Mpc in length.  Given the linear arrangement of galaxies along 
Virgo's principal axis,  it is natural to ask whether this might be 
related to filamentary features on larger scales. 

As viewed on the plane of the sky, Virgo's principal axis points in 
the direction of Abell 1367, a rich cluster located some 
$75\,h_{75}^{-1}$ Mpc away along a projected position angle of 
$125^{\circ}$.  A1367 itself forms part of a well-known supercluster 
with the Coma cluster \citep{gt78,del86}.  This raises the intriguing 
possibility that the Virgo, A1367 and Coma clusters may all 
be members of a common filamentary network, an idea 
suggested two decades ago by \citet{zel82}.  

Figure 4 plots the distribution of nearby poor galaxy clusters 
from the catalog of \citet{white99}.  A narrow bridge of material 
is clearly seen connecting Virgo and A1367.   Furthermore, 
the chain of giant elliptical galaxies that defines Virgo's principal
axis appears to be a segment of this filament.  Hence, the Virgo cluster
points towards A1367, not only in two dimensions, but in three.
Additional indirect evidence of a Virgo-A1367 connection comes
from A1367's orientation; X-ray observations show that 
this cluster is very elongated along a projected position angle 
of $\sim 140^{\circ}$ \citep{jf99}, and thus in the general direction 
of the Virgo cluster.  

Of course, the Virgo cluster is often considered part of the larger
supercluster which includes Hydra-Centaurus and Pavo-Indus (e.g., Tully 1986).
\citet{lb88} viewed this extended, planar supercluster complex
as centered on the ``Great Attractor,'' with Virgo near the outskirts.
Thus, we can now trace an apparent link between two of the most massive
structures in the local universe: from the ``Great Wall'' encompassing
the Coma cluster to the Great Attractor, through A1367 and the Virgo cluster.

\section{Conclusion}

We have shown that the brightest elliptical galaxies in the Virgo 
cluster have a remarkably collinear arrangement in three dimensions.    
This axis appears to be part of a larger filament connecting the Virgo 
cluster to Abell 1367.  Virgo's elliptical galaxies also exhibit a 
tendency for their major axes to share this same orientation.

Cosmological N-body simulations show that clusters of galaxies often 
form at the intersection of filaments, with material flowing into the 
cluster along one or more axes \citep{vhaarlem93,bond96}.  
Built by a series of subcluster mergers that occur along preferred 
directions, clusters naturally develop major axis orientations that 
reflect the orientation of the dominant filament feeding them 
\citep{wjf95}.  Furthermore, if large elliptical galaxies are  
products of galaxy mergers, then the highly anisotropic nature 
of the merger process will tend to produce ellipticals 
whose major axis orientations are also aligned with the 
surrounding filamentary structure \citep{west94,dub98}.

The results presented here are consistent with a picture 
in which the formation of the Virgo cluster and its elliptical 
galaxy population has been driven by anisotropic 
inflow of material along the Virgo-A1367 filament.
Although Virgo may be fed by more than one filament \citep{tully82},
the one joining the cluster to A1367 appears to dominate. 
Additional evidence in support of this interpretation comes 
from X-ray observations of M86, which show a plume of hot gas 
being stripped from this galaxy along a projected position angle 
of $\sim 110^{\circ}$ \citep{forman79}, presumably as a result of 
ram pressure as the galaxy travels along the cluster principal axis
at over 1200~km$^{-1}\,$s.

Finally, it is intriguing that M87's famous jet also emanates
along the same direction as the Virgo cluster major axis.  
This is true not only in two dimensions, where both are oriented 
along projected position angles of $\sim 110-120^{\circ}$, but also 
in three dimensions.
Detailed models of the jet's observed properties  
indicate that it is most likely oriented within $\sim 20$ degrees of
the line of sight \citep{heinz97,biretta99}, very close to the 
$10^{\circ}$ to $15^{\circ}$ line of sight inclination angle 
of the cluster principal axis found in \S 2.
While this might be purely coincidental, alternatively, it might 
be an indication that Virgo's principal axis has influenced not 
only the orientations of its member elliptical galaxies, but 
perhaps even the massive blackhole at the center of M87 that is 
believed to power its nuclear activity \citep{west94}.

\acknowledgments

We are grateful to John Tonry, Alan Dressler, and Ed Ajhar for allowing us 
access to the SBF Survey data prior to publication.
MJW acknowledges support from NSF grant 007114, and from NSERC of Canada.

\clearpage

\begin{figure}
\plotone{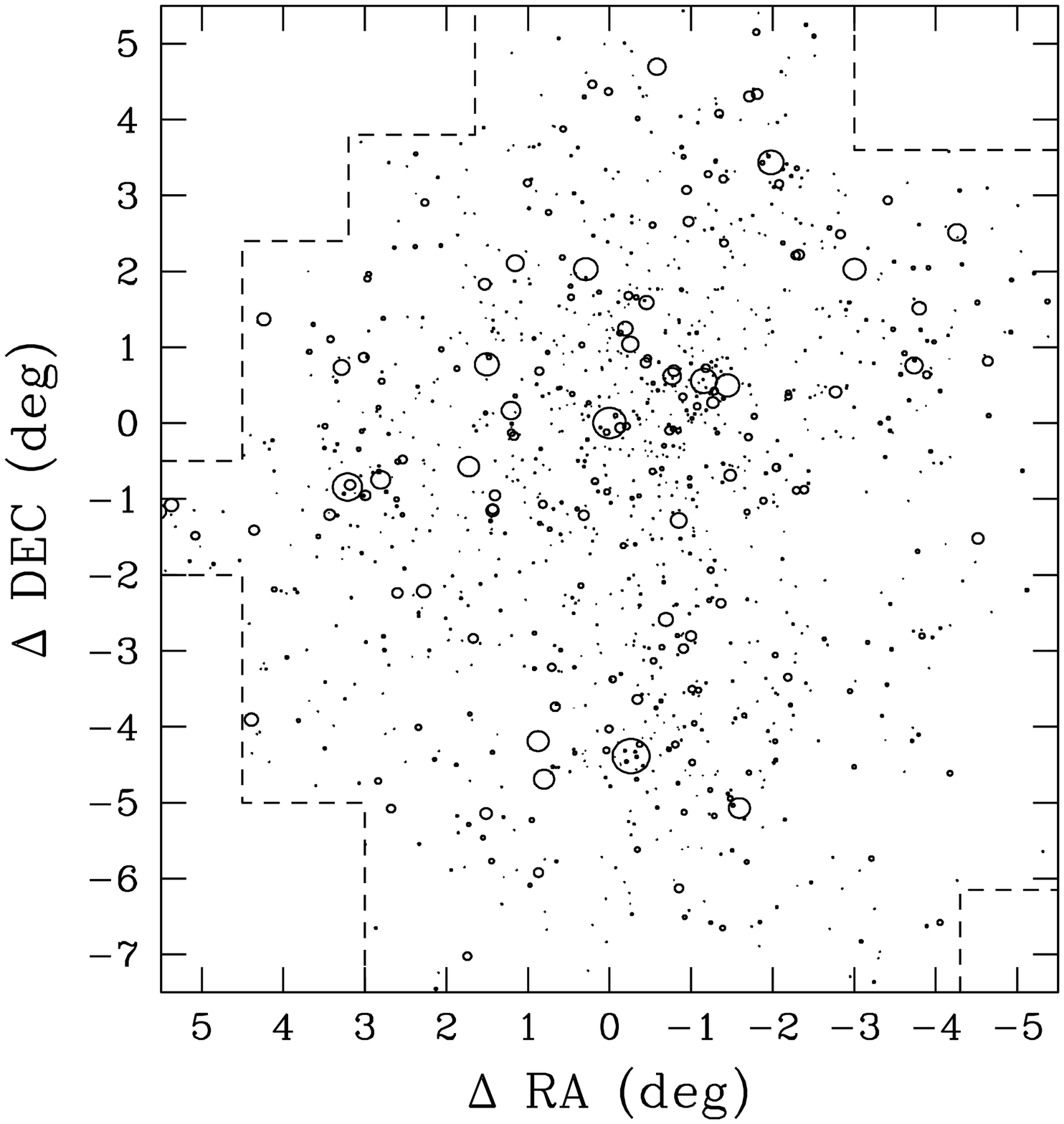}
\caption{A map of the distribution of likely member 
galaxies in the Virgo cluster, taken from the catalogue 
of \cite{bing85}.  
Galaxy positions are measured relative to NGC 4886 (M87).  
The area of each symbol is proportional to galaxy luminosity.  The dashed 
lines indicate regions not included in the Binggeli et al. survey.  
As discussed in the text, the brightest galaxies in the northern part of 
the cluster form a chain oriented along a projected position angle of 
$\sim 110$ degrees.  
}
\end{figure}

\begin{figure}
\plotone{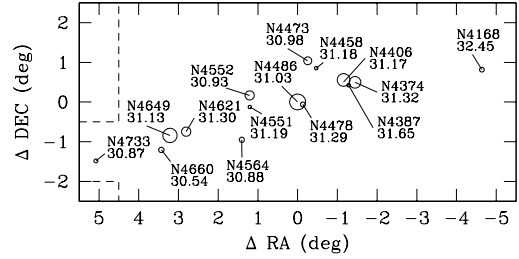}
\caption{A plot of those elliptical galaxies in the northern half of
the Virgo cluster for which SBF 
distances are available from the survey of \citet{tonry00}.
We have also included NGC 4168 which, although not a Virgo 
member, lies in the field of view and also has a measured SBF distance. 
}
\end{figure}

\begin{figure}
\plotone{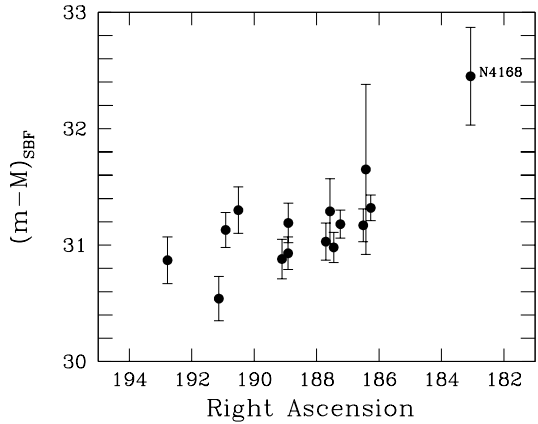}
\caption{Right ascension versus SBF distance modulus for the 
Virgo ellipticals in Table 1.  NGC 4168 is also indicated.}
\end{figure}

\begin{figure}
\plotone{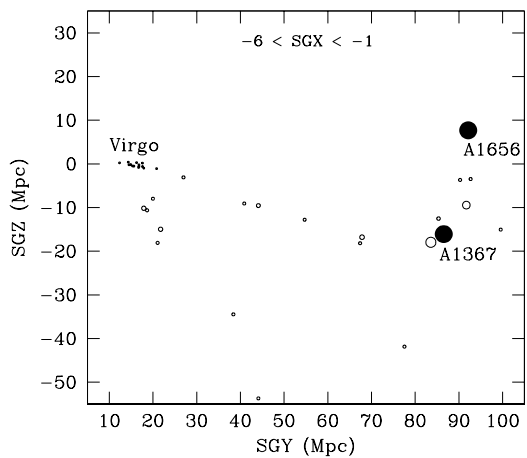}
\caption{The distribution of poor clusters in a $5\,h_{75}^{-1}$ Mpc thin 
slice centered on the Virgo cluster.  Supergalactic coordinates SGY and SGZ 
are plotted.  The poor cluster sample is taken from \citet{white99}.  
Cluster distances are based on the mean redshift of all members.  Only 
clusters with $\langle z \rangle \geq 0.005$ were used, to 
ensure that peculiar velocities are small enough relative to Hubble 
expansion that accurate distances could be obtained from the Hubble law. 
The area of each symbol is proportional to the number of galaxies belonging 
to the group.  Virgo ellipticals listed in Table 1 are also plotted as 
individual points.  A narrow bridge of material is seen connecting the 
Virgo cluster with A1367.
}
\end{figure}


\clearpage

\begin{deluxetable}{lcr}
\tablecaption{Virgo Cluster Ellipticals. \label{tbl-1}}
\tablewidth{0pt}
\tablehead{
\colhead{Galaxy} & \colhead{$(m-M)_{SBF}$}   & \colhead{PA (deg)} 
}
\startdata
NGC 4374 (M84) &$31.32 \pm 0.11$  & 135 \\
NGC 4387 &$31.65 \pm 0.73$  & 140 \\
NGC 4406 (M86) &$31.17 \pm 0.14$  & 130 \\
NGC 4458 &$31.18 \pm 0.12$  & 46 \\
NGC 4473 &$30.98 \pm 0.13$  & 100 \\
NGC 4478 &$31.29 \pm 0.28$  & 140 \\
NGC 4486 (M87) &$31.03 \pm 0.16$  & 160 \\
NGC 4551 &$31.19 \pm 0.17$  &  70 \\
NGC 4552 (M89) &$30.93 \pm 0.14$  & 125{\tablenotemark{a}}\\
NGC 4564 &$30.88 \pm 0.17$  &  47 \\
NGC 4621 (M59) &$31.31 \pm 0.20$  & 165 \\
NGC 4649 (M60) &$31.13 \pm 0.15$  & 105 \\
NGC 4660 &$30.54 \pm 0.19$  & 100 \\
NGC 4733 &$30.87 \pm 0.20$  & 115 \\
\enddata
\tablenotetext{a}{With the exception of NGC4552, all 
major axis position angles are taken from the Lyon-Meudon Extragalactic Database 
(LEDA).  No PA was listed for NGC4552 in LEDA, and so we 
adopted a value of $125^{\circ}$ from the study by 
Carollo et al. (1997; see also King 1978).
}
\end{deluxetable}


\newpage
\begin{thebibliography}{}
\bibitem[Arp(1968)]{arp68} Arp, H. 1968, \pasp, 80, 129 
\bibitem[Binggeli(1982)]{bing82} Binggeli, B. 1982, \aap, 107, 338
\bibitem[Binggeli, Sandage \& Tammann(1985)]{bing85} Binggeli, B., 
Sandage, A., \& Tammann, G.A. 1985, \aj, 90, 1681
\bibitem[Binggeli, Tammann \& Sandage(1987)]{bing87} Binggeli, B.,  
Tammann, G.A., \& Sandage, A. 1987, \aj, 94, 251
\bibitem[Binggeli, Popescu \& Tammann(1993)]{bing93} Binggeli, B., Popescu, C.C., 
\& Tammann, G.A. 1993, \aaps, 98, 275
\bibitem[Binggeli(1999)]{bing99} Binggeli, B. 1999, in The Radio Galaxy Messier 87,
Ringberg Workshop, eds. H.-J. R\"oser \& K. Meisenheimer, Springer, Berlin, 9
\bibitem[Biretta, Sparks \& Macchetto(1999)]{biretta99} Biretta, J.A., 
Sparks, W.B., \& Macchetto, F. 1999, \apj, 520, 621
\bibitem[Blakeslee, Ahjar \& Tonry(1999)]{bat99} Blakeslee, J.P., Ajhar, E.A., 
\& Tonry, J.L. 1999, in Post-Hipparcos Cosmic Candles, eds. A. Heck \& 
F. Capulo (Boston: Kluwer), 181
\bibitem[Blakeslee, Vazdekis, \& Ajhar(2000)]{blakeslee00} Blakeslee, J.P., 
Vazdekis, A., \& Ajhar, E.A. 2000, \mnras, in press
\bibitem[B\"ohringer et al.(1994)]{bohr94} B\"ohringer, H.,  Briel, U.G.,  
Schwarz, R.A., Voges, W., Hartner, G. \& Tr\"umper, J. 1994, \nat, 368, 828
\bibitem[Bond, Kofman \& Pogosyan(1996)]{bond96} Bond, J.R., Kofman, L. \& 
Pogosyan, D. 1996, \nat, 380, 603
\bibitem[Carollo et al.(1997)]{carollo97} Carollo, C.M., Franx, M., 
Illingworth, G.D., \& Forbes, D.A. 1997, \apj, 481, 710
\bibitem[de Lapparent, Geller \& Huchra(1986)]{del86} de Lapparent, V., 
Geller, M.J., \& Huchra, J.P. 1986, \apj, 302, L1
\bibitem[Dubinski(1998)]{dub98} Dubinski, J. 1998, \apj, 502, 141
\bibitem[Ferrarese et al.(2000)]{ferrarese00} Ferrarese, L., et al.\ 
2000, \apj, 529, 745 
\bibitem[Forman et al.(1979)]{forman79} Forman, W., Schwarz, J., 
Jones, C., Liller, W., \& Fabian, A.C. 1979, \apj, 234, L27
\bibitem[Heinz \& Begelman(1997)]{heinz97} Heinz, S. \& Begelman, M.C. 1997, 
\apj, 490, 653
\bibitem[Jacoby et al. (1992)]{jacoby92} Jacoby, G.H., et al. 1992, \pasp, 
104, 599
\bibitem[Jones \& Forman(1999)]{jf99} Jones. C. \& Forman, W. 1999, \apj 
511, 65
\bibitem[King(1978)]{king78} King, I.R. 1978, \apj, 222, 1
\bibitem[Liu et al.(2000)]{liu00} Liu, M. C., Charlot, S., \& Graham, 
J. R. 2000, \apj, in press
\bibitem[Lynden-Bell et al.(1988)]{lb88} Lynden-Bell, D., Faber, S. M., Burstein, D., Davies, R. L., Dressler, A.,
 Terlevich, R. J., \& Wegner, G. 1988, \apjl, 326, 19
\bibitem[Neilsen and Tsvetanov(2000)]{neilsen00} Neilsen, 
E.H., Jr. \& Tsvetanov, Z.I. 2000, \apj, 536, 255 
\bibitem[Porter, Schneider \& Hoessel(1991)]{porter91} Porter, A.C., 
Schneider, D.P., \& Hoessel, J.G. 1991, \aj, 101, 1561 
\bibitem[Schindler, Binggeli \& B\"ohringer(1999)]{schind99} Schindler, S., 
Binggeli, B., \& B\"ohringer, H. 1999, \aap, 343, 420
\bibitem[Gregory \& Thompson(1978)]{gt78} Gregory, S.A., \& Thompson, 
L.A. 1978, \apj, 222, 784
\bibitem[Tonry, Ajhar \& Luppino(1990)]{tonry90} Tonry, J.L., Ajhar, E.A., 
\& Luppino, G.A. 1990, \aj, 100, 1416
\bibitem[Tonry et al.(1997)]{tonry97} Tonry, J.L., Blakeslee, J.P.,
Ajhar, E.A., \& Dressler, A. 1997, \apj, 475, 399
\bibitem[Tonry et al.(2000)]{tonry00} Tonry, J.L., Dressler, A., 
Blakeslee, J.P., Ajhar, E.A., Fletcher, A.B., Luppino, G.A., 
Metzger, M.R., \& Moore, C.P. 2000, \apj, in press
\bibitem[Tonry \& Schneider(1988)]{tonry88} Tonry, J., \& Schneider, 
D.P. 1988, \aj, 96, 807
\bibitem[Tully(1982)]{tully82} Tully, R.B. 1982, \apj, 257, 389
\bibitem[Tully(1986)]{tully86} Tully, R.B. 2986, \apj, 303, 25
\bibitem[van Haarlem \& van de Weygaert(1993)]{vhaarlem93} van Haarlem, M. \& 
van de Weygaert, R. 1993, \apj, 418, 544
\bibitem[West(1994)]{west94} West, M.J. 1994, \mnras, 268, 79
\bibitem[West(1998)]{west98} West, M.J. 1998, A New Vision of an Old Cluster:
 Untangling Coma Berenices, A. Mazure et al., Singapore: World Scientific, 36
\bibitem[West, Jones \& Forman(1995)]{wjf95} West, M.J., Jones, C., \& 
Forman, W. 1995, \apj, 451, L5
\bibitem[Weil et al.(1997)]{weil97} Weil, M., Bland-Hawthorn, J., \& 
Malin, D.F. 1997, \apj, 490, 664
\bibitem[White et al.(1999)]{white99} White, R.A., et al. 1999, \aj, 118, 
2014
\bibitem[Worthey(1994)]{worthey94} Worthey, G. 1994, \apjs, 95, 107
\bibitem[Young \& Currie(1998)]{yc98} Young, C.K., \& Currie, M.J. 1998, 
\aaps, 127, 367
\bibitem[Zeldovich, Einasto \& Shandarin(1982)]{zel82} Zel'dovich, Ya.B., 
Einasto, J., \& Shandarin, S.F. 1982, \nat, 300, 407
\end{thebibliography}
\end{document}